\documentclass[11pt]{article}
\usepackage[utf8]{inputenc} 
\usepackage{amsmath,amsthm,amsfonts,amssymb,amscd}
\usepackage{booktabs}
\usepackage[usenames,dvipsnames]{color} 
\usepackage{color}
\usepackage{fullpage}
\usepackage{lastpage}{}
\usepackage{enumitem}
\usepackage{fancyhdr}
\usepackage{mathrsfs}
\usepackage{wrapfig}
\usepackage{setspace}
\usepackage{calc}
\usepackage{multicol}
\usepackage{cancel}
\usepackage[retainorgcmds]{IEEEtrantools}
\usepackage[margin=3cm]{geometry}
\usepackage{amsmath}

\usepackage{empheq}
\usepackage{framed}
\usepackage[most]{tcolorbox}
\usepackage{xcolor}
\usepackage{verbatim}
\usepackage{authblk}
\usepackage[colorlinks=true,
            linkcolor=red,
            urlcolor=blue,
            citecolor=blue]{hyperref}
\usepackage{cite}
\usepackage{pdfpages}
\colorlet{shadecolor}{orange!15}
\theoremstyle{definition}

\begin{document}\date{\vspace{-8ex}}
\author[1,2]{Changhao Li} \author[1,3]{Mo Chen} \author[1]{Dominika Lyzwa} \author[1,2]{Paola Cappellaro \thanks{pcappell@mit.edu}}
\affil[1]{Research Laboratory of Electronics, Massachusetts Institute of Technology, Cambridge, MA 02139}
\affil[2]{Department of Nuclear Science and Engineering,Massachusetts Institute of Technology, Cambridge, MA 02139}
\affil[3]{Department of Mechanical Engineering, Massachusetts Institute of Technology, Cambridge, MA 02139}
\renewcommand\Affilfont{\fontsize{10.2}{10.6}\itshape}
\title{\LARGE \bf All-optical quantum sensing of rotational Brownian motion of magnetic molecules}
\maketitle
\thispagestyle{empty}

 \begin{abstract}
Sensing local environment through the motional response of small molecules lays the foundation of many fundamental technologies.
 The information of local viscosity, for example, is contained in the random rotational Brownian motions of molecules.
 However, detection of the motions is challenging for molecules with sub-nanometer scale or high motional rates. 
Here we propose and experimentally demonstrate a novel method of detecting fast rotational Brownian motions of small magnetic molecules. 
With  electronic spins as sensors, we are able to detect changes in motional rates, which yield different noise spectra and therefore different relaxation signals of the sensors.
As a proof-of-principle demonstration, we experimentally implemented this method to detect the motions of gadolinium (Gd) complex molecules with nitrogen-vacancy (NV) centers in nanodiamonds. 
With all-optical measurements of the NV centers' longitudinal relaxation, we distinguished binary solutions with varying viscosities.
Our method paves a new way for detecting fast motions of sub-nanometer sized magnetic molecules with better spatial resolution than conventional optical methods. It also provides a new tool in designing better contrast agents in magnetic resonance imaging.
 \end{abstract}

\section{Introduction.}

Characterizing local environments and capturing  dynamical variations of local quantities can provide important information about physical and biological processes. In particular, detecting  variations in molecules' Brownian motions would yield information about particle size and local viscosity in biological sensing~\cite{BioNanoBook2007}. For example, the random rotations of small molecules can respond to environmental changes and reveal local dynamics and biological functions. 
Conventional optical microscopy methods are capable of measuring  the rotational Brownian motions (RBM) of particles in the sub-micron scale or larger, using dark-field microscopy~\cite{DarkField1,DarkField2},   microrheological techniques based on particle's anisotropy~\cite{Anisotropy1,Anisotropy2} or fluorescence polarization spectroscopy~\cite{Fpolarization1}.  However, it is very difficult to extend these techniques to the nanometer regime:
 the rotational motion of nanometer-size molecules is typically much smaller than the optical diffraction limit, therefore it cannot be directly captured; in addition, their motional rates are typically in the GHz range, which are beyond the detection rates of most optical techniques. 
To this end, an all-optical method capable of capturing the fast rotations is still lacking. 

Quantum sensors~\cite{Degen17} have recently emerged as a powerful tool to explore properties at the nanoscale. 
For example, nitrogen-vacancy (NV) color centers in diamond have shown the potential to study magnetic~\cite{JorgNC2013,Ania2014PRApp,AjoyPRX2015,Jayich2018NC} and electric~\cite{Dolde11} fields, temperature~\cite{Kucsko13,Toyli13,JorgtempNL2013,LiuPRXNDthermal}, strain~\cite{FuchsPRL2013,Teissier14,Udvarhelyi18} and other quantities, displaying an exceptional sensitivity and spatial resolution even at room temperature,  thanks to the control on their spin states.
In particular, NV centers in nanodiamonds (NDs) have been employed to sense various properties~\cite{HollenbergPNAS2013,JacquesPRB2013,JorgNC2017,LiuNC_Hydrogel} that affect, and reveal, chemical and biological processes. NV centers in NDs have many favorable properties, ranging from very high photo- and thermal-stability, to bio-compatibility. All these desirable properties make NDs preferable over organic dyes in biosensing applications~\cite{Chang08}. 
Furthermore, the size of NDs containing stable NV centers  can reach down to just several nanometers~\cite{Tisler09}, promising nanometer spatial resolution. 

Here we propose a novel method that exploits the electronic spins of NV centers in NDs to detect the fast RBMs of sub-nanometer sized magnetic molecules, overcoming the limitations of conventional optical approaches. The key idea is that fast RBM rates modify the spectrum of the magnetic noise generated by the magnetic molecules, slowing down the relaxation of our spin sensors in a way analogous to motional narrowing in NMR~\cite{NMRbook1983}. The all-optical measurement of the relaxation rates of our spin sensors captures the RBMs of the target magnetic molecules, while providing a simple and accessible experimental technique. We point out that while we consider RBM of magnetic molecules here, this method can be extended to detect other molecules' RBM by attaching magnetic labels to them.

We demonstrate this novel method in a proof-of-principle experiment. The longitudinal relaxation of NV centers in NDs is monitored in the presence of gadolinium (Gd) complex molecules, which are among the most commonly used contrast agents in magnetic resonance imaging (MRI). We observed a difference in the relaxation rate  corresponding to changes in the RBM rate when varying the local viscosity, thus demonstrating the feasibility of the proposed method.

Conventionally, the rotational motion of Gd(III) complex molecules can be indirectly extracted from EPR spectra with nuclear magnetic relaxation dispersion (NMRD) analysis~\cite{Rast2000,Rast2001,MerbachBook2013}. However, this technique not only entails demanding experimental conditions, but also it
does not provide the complete picture of Gd molecules RBM. 
For example, the rotational correlation time obtained from proton's or $^{17}$O's relaxation measurements represents, instead of the rotation information for the whole molecule, the local rotational correlation time of specific bond vectors. 
Compared with the NMRD technique, our method is more robust as it measures the effective rotational motion rates of the whole molecule.  Moreover, in principle, the spatial resolution is only limited by ND size, which is in the nanometer regime. The experimental apparatus is simpler, as NV centers  allow all-optical initialization and readout, without the need for a strong external magnetic field.  Finally, our method is fully compatible with microfluidic techniques, supporting study of RBM in living cells where sample volume is constrained. 
Our technique  would then provide a versatile tool to understand the dynamics of magnetic molecules and design better contrast agents in (functional) MRI~\cite{MerbachBook2013,MoleculefMRI,CaSensor}. For instance, magnetic molecules with long rotational correlation times are usually preferred for higher relaxivity. 
Biologically relevant ligand coatings or bindings in synthesized magnetic nanoparticles could increase the particles' size and therefore increase their relaxivity.
The method we propose here provides an independent study of rotational correlation time and would be helpful in quantifying the performance of contrast agents.

\begin{figure}[ht]
\centering
\includegraphics[scale=0.34]{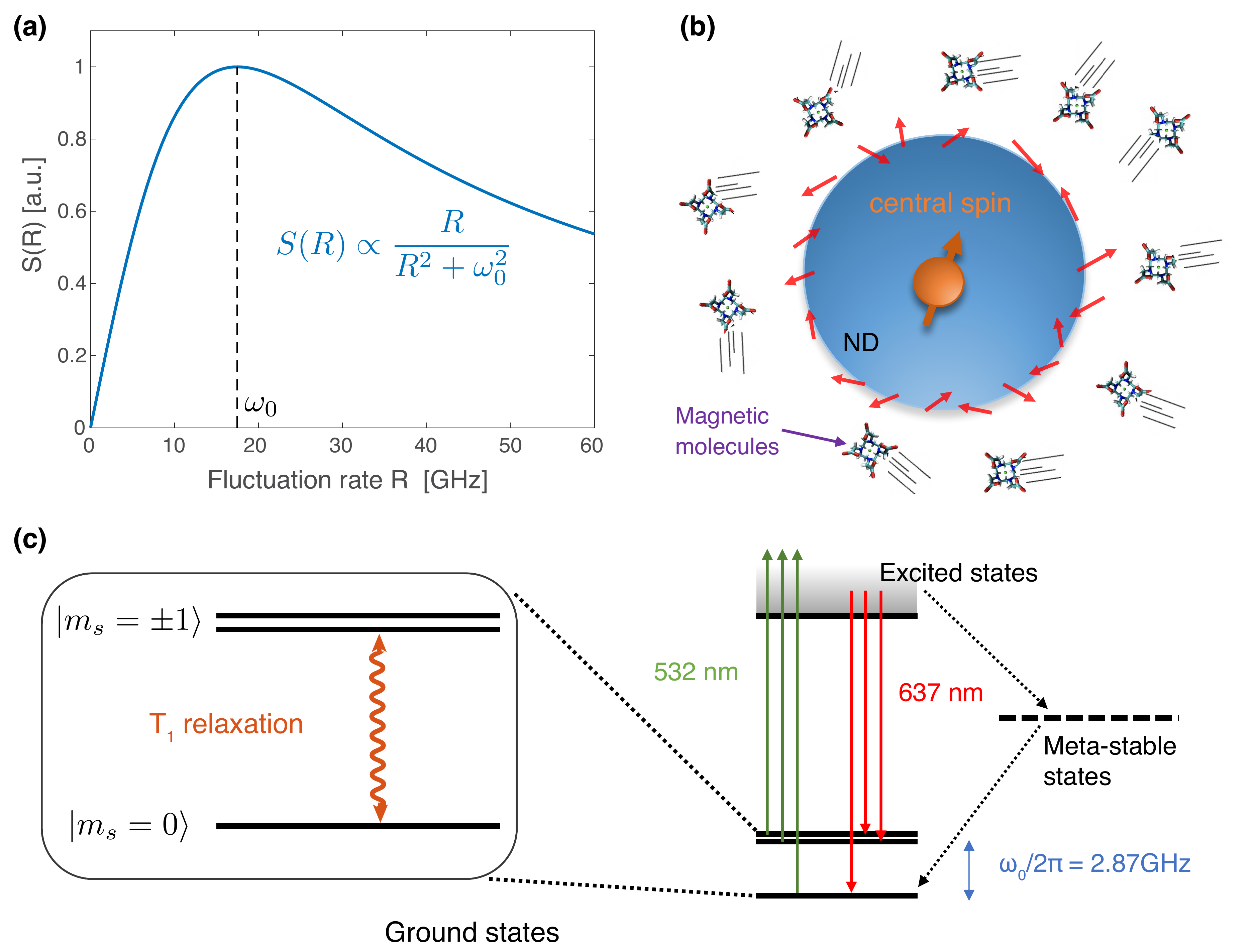}
\caption{\label{Fig1} \textbf{(a)} Lorentzian spectrum as a function of noise fluctuation rate.  \textbf{(b)} Electronic spin (for example, a NV center) in a host crystal such as diamond, are subjected to the noise generated by  surface spins (red arrows) and surrounding magnetic complexes (such as Gd-DOTA) with random translational and rotational Brownian motions.   \textbf{(c)} Right: Energy level diagram of an NV center showing the optical transitions. Green laser non-resonantly excites NV spin to the excited state,  and the NV decays back emitting red fluorescence photons. The $|m_{s}= \pm1\rangle $ states can also decay non-radiatively through the meta-stable singlet state and back to $|m_{s} = 0\rangle$ ground state during excitation (dashed lines), providing a mechanism for both optical initialization to $|m_s=0\rangle$ and spin state-dependent optical readout. Left: spin polarized in one of the ground states will reach thermal equilibrium with a time scale T$_{1}$, known as longitudinal relaxation process. The presence of the external transverse magnetic noise will accelerate this process.}
\end{figure} 

\section{Theoretical principle}

The fast RBMs of small magnetic particles contribute to high frequency fluctuations in the magnetic noise spectrum. This effect can be detected by nearby spins, whose relaxation times are sensitive to external magnetic noise. Counterintuitively, the added motion actually often lengthens the spin relaxation time.  To understand how,  consider a Lorentzian spectrum as a function of the noise fluctuation rate $R$ (Fig.~\ref{Fig1}.(a)). When the fluctuation rate is larger than the resonant frequency of the spin sensors $R>\omega_0$, a further increase in fluctuation rate results in a smaller noise spectrum intensity at the spin resonance frequency. This effect is similar to motional narrowing in NMR~\cite{NMRbook1983}, and leads to a longer relaxation time of the spin sensors.

To provide a more quantitative description, we analyze all the effects contributing to the longitudinal relaxation time T$_{1}$. In the presence of magnetic particles, including paramagnetic impurities  on the surface and  in the surrounding solution, the  T$_{1}$ time of a single NV electronic spin  is given by~\cite{JacquesPRB2013}:
\begin{eqnarray}\label{eq:fermigoldenrule}
\frac{1}{T_{1}} = \frac{1}{T_{1,bulk}} + \sum_{k} 3\gamma_{k}^{2}B^{2}_{\perp,k} \frac{\tau_{c,k}}{1+\omega_{0}^{2}\tau_{c,k}^{2}},
\end{eqnarray}
where $T_{1,bulk}$~is the relaxation time of NV in bulk diamond and $\omega_{0}=(2\pi)2.87$~GHz  the NV energy level splitting between the $|m_s=0\rangle$ and $|m_s=\pm1 \rangle$ states at zero external magnetic field.
For each magnetic particle $k$,  $\gamma_k$ is the gyromagnetic ratio; $B_{\perp,k}$ is the rms transverse magnetic noise strength (see the supplementary material~\cite{SI});  and $\tau_{c,k}=1/R_{k}$ is the noise correlation time, the inverse of the noise fluctuation rate. 
As already mentioned, when $R_{k} > \omega_{0}$, any further increase in $R_{k}$ results in longer T$_{1}$. 

As an example, we consider using T$_1$ relaxation to detect the RBM of Gd(III) chelators. 
The total noise fluctuation rate from Gd(III) molecules $R_{Gd,tot}$~is given by~\cite{JorgNC2013}:
\begin{eqnarray}
R_{Gd,tot}=R_{Gd,dip}+R_{vib}+R_{trans}+R_{rot}
\end{eqnarray}
where $R_{Gd,dip}$ represents the dipolar interaction rates between Gd molecules, $R_{vib}$~the intrinsic vibrational rate between the Gd ion's vibrational energy levels, and $R_{trans (rot)}$~the translational (rotational) Brownian motion rates.  
Variations in the fluctuation rate $R_{Gd,tot}$  induce changes in the T$_{1}$ signal decay time of NV electronic spins via Eq.~(\ref{eq:fermigoldenrule}). It is then possible to detect RBM and sense the local environment through T$_1$~relaxometry of NV sensors in an all-optical fashion.

\section{Experimental demonstration}

\begin{figure}[h!]
\centering
\includegraphics[scale=0.34]{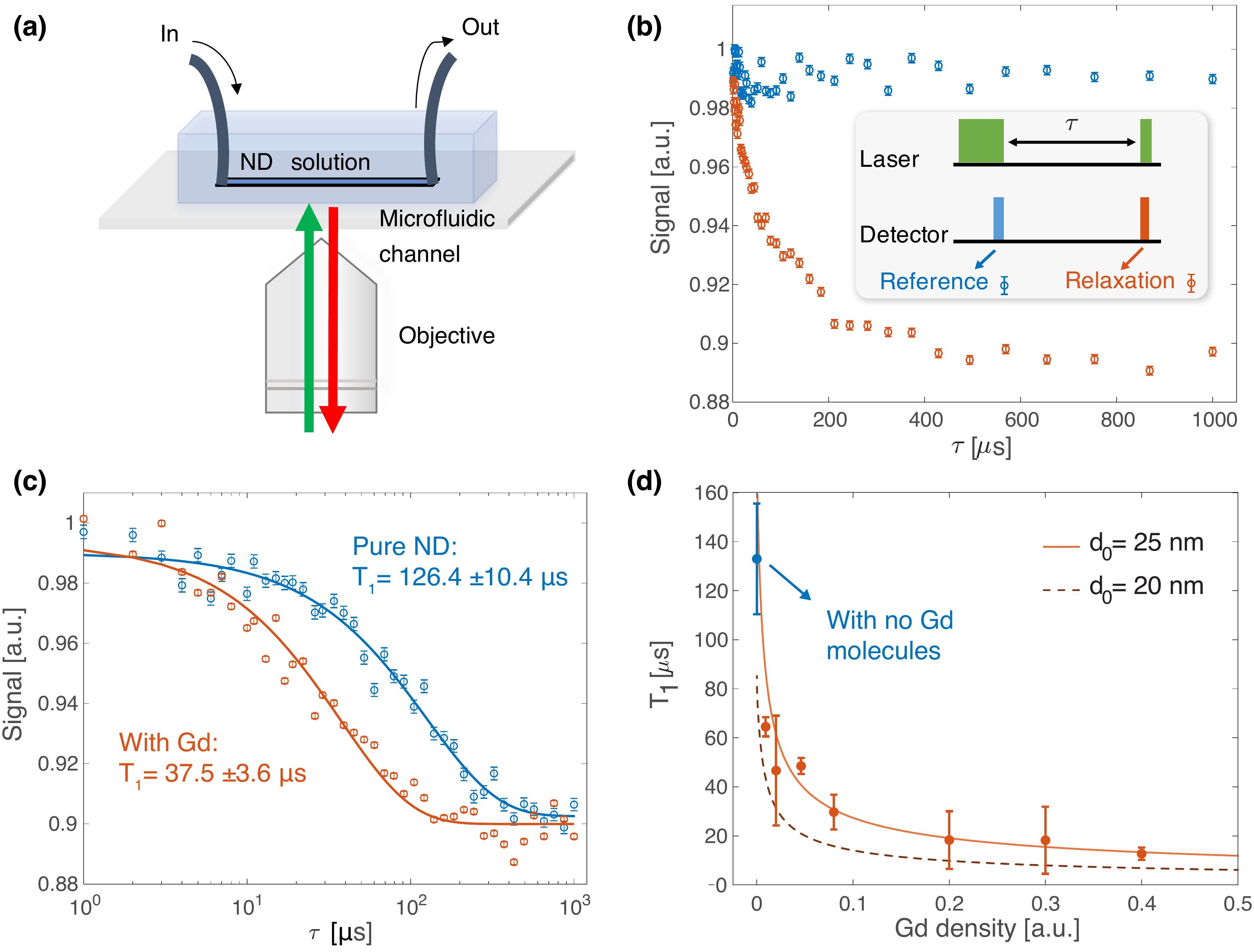}
\caption{\label{Fig2} \textbf{(a) }Diagram of the experimental setup. A nanodiamond (ND) solution with Gd molecules was loaded into a microfludic channel. We excited the ND sample with green laser and collected the red fluorescence with the same objective. \textbf{(b)} Typical T$_{1}$ measurement with reference and relaxation signal lines. The error bars for each data point are the standard deviations over repetitions.  
In the inset:  T$_{1}$ relaxometry protocol.  
\textbf{(c)} T$_{1}$ quenching in the presence of Gd molecules. The experimental data shows the relaxation signal normalized by the reference, in the presence (red) and absence (blue) of Gd. We fit the decay to a single exponential (solid curves) and use the fitting error for the error bars of the data.
 \textbf{(d)} Relaxation time T$_{1}$  as a function of the density of Gd molecules (red). The bare ND relaxation time (blue) is around 130 $\mu s$.
We measured the relaxation signal (as in (c)) over several spatially separated spots and took the average of the fitted T$_1$. 
  Error bars  are the propagated fitting errors from the different spots. The experimental data matches the theoretical predictions with ND sizes of $20$~nm (dashed lines) and $25$~nm (solid lines), assuming higher Gd density than the prepared solution, due to aggregation (see main text).}
\end{figure}

\begin{figure}[htb]
\centering
\includegraphics[scale=0.34]{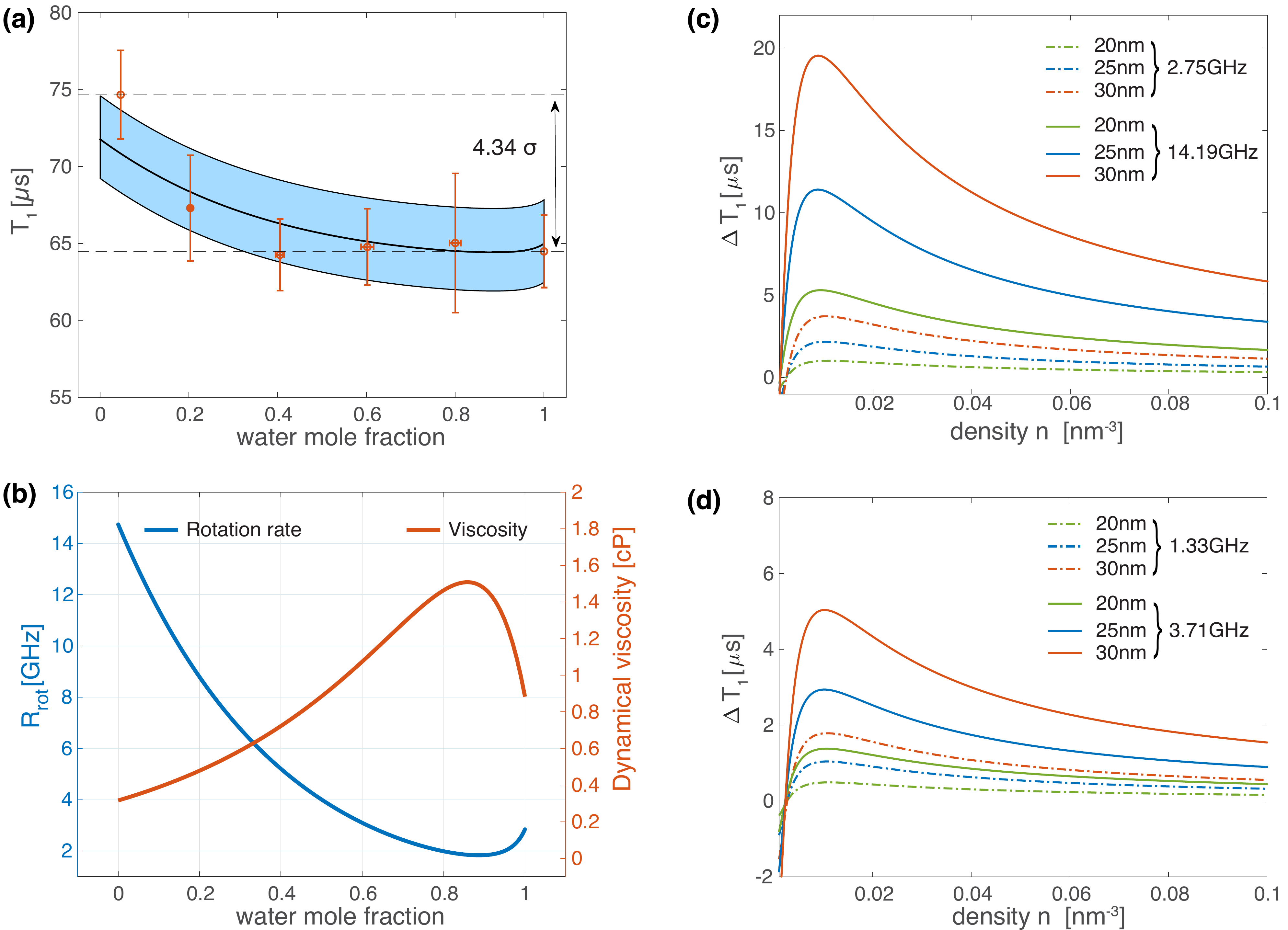}
\caption{\label{Fig3}  
\textbf{(a)} Experimental measurement of the relaxation time  in different water-acetone binary solvents (the lowest water mole fraction we prepared was 0.046). The error bars are propagated errors from ensemble measurements over different confocal spots. The solid line corresponds to the theoretical estimation where the Gd molecule density is obtained by approximately matching the experimental result for (mean) T$_1$ in pure water with the theoretical prediction. The shaded area corresponds to $10 \%$ error for the Gd density.
\textbf{(b)}  Predicted dynamical viscosity as well as motional rates $R_{rot}$ as a function of water mole fraction in water-acetone binary solutions.
\textbf{(c-d)} Increase in relaxation time T$_1$ due to rotational RBM over the T$_1$ obtained in the absence of any rotational motion contribution. The rotational RBM rate, and thus the T$_1$ depend on the Gd molecules density, on the viscosity (solid line for acetone and dashed lines for water), and on the ND diameter. The change in relaxation, $\Delta$T$_1$ is more marked when we consider the  microviscosity factor $f_{r}$ (c),  while neglecting it (d) the predictions do not match our experimental results.
}
\end{figure}

\begin{figure}[htb]
\centering
\includegraphics[scale=0.34]{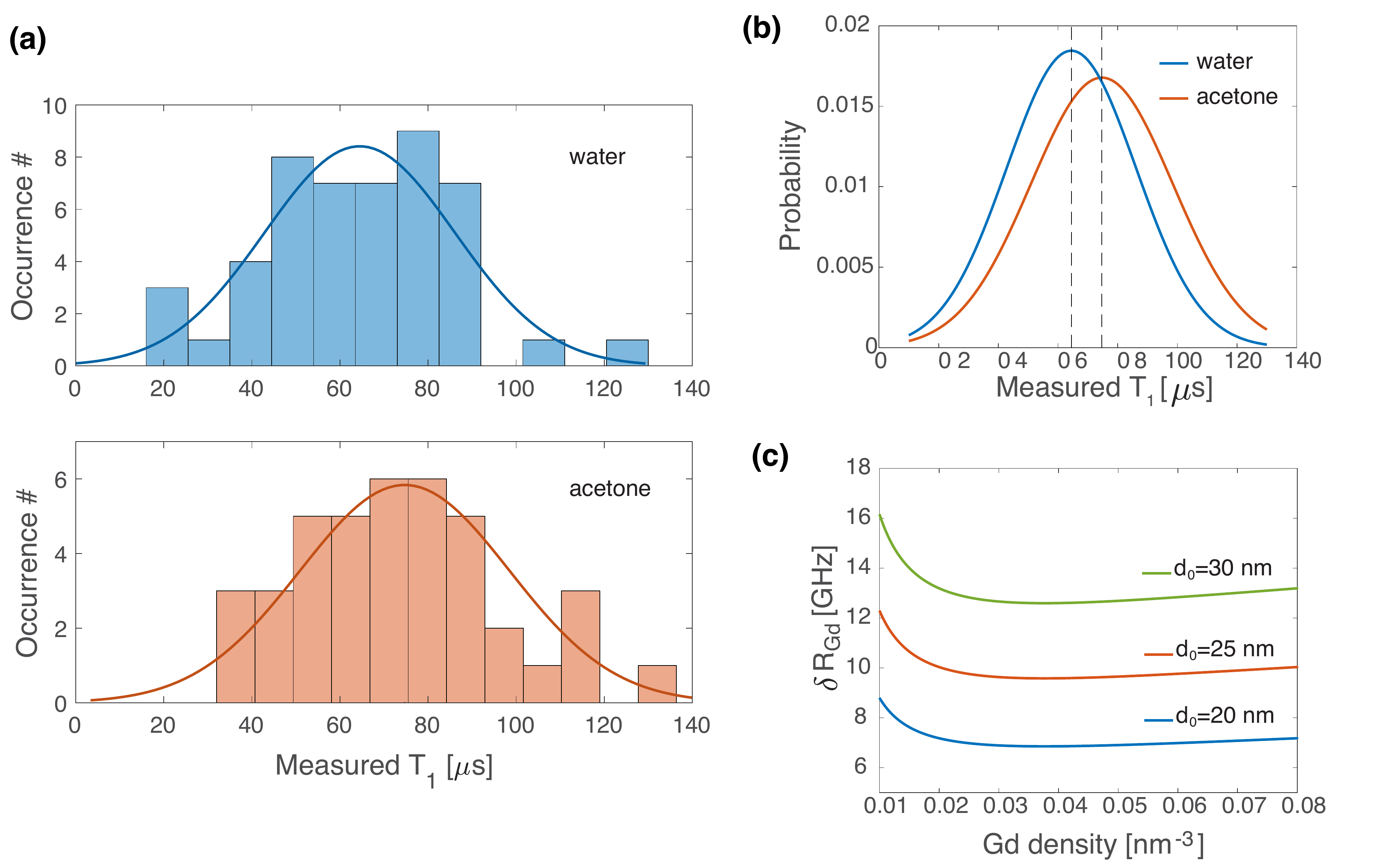}
\caption{\label{Fig4} \textbf{(a)} Distribution of measured relaxation times over different confocal points for molecules in (top) pure water   and (bottom) nearly pure acetone (with water mole fraction 0.046). \textbf{(b)}  Gaussian fittings to the two distributions show two clearly distinct peak values. \textbf{(c)} Sensitivity estimation as a function of Gd molecule density for a single NV sensor located at the center of a nanodiamond with diameter d$_{0}$ in a 10 s data acquisition time.}
\end{figure}

Next, we demonstrate our proposed method in a proof-of-principle experiment. We use NDs hosting NV centers as sensors to detect the RBM rates of one type of Gd(III) complex, gadolinium 1,4,7,10-tetraazacyclododecane-N,N',N'',N'''-tetraacetate (Gd-DOTA). 
At zero magnetic field, the NV center has an electronic spin-1 ground state with resonance frequency $\omega_{0}=(2\pi)2.87$~GHz (Fig.~\ref{Fig1}(c)). 
The spin can be optically polarized and read out with $532$~nm laser excitation. 
In the absence of laser illumination, the spin state population will reach its thermal equilibrium with a time scale T$_{1}$.
 
We measured this longitudinal relaxation time T$_{1}$ of an ensemble of NV centers with a home-built confocal microscope at room temperature, without any applied external magnetic field. The NDs we used have an average size of around $25$~nm and are terminated with carboxyl groups. In the absence of Gd-DOTA, we measured a relaxation time of about $130~\mu$s, which is significantly shorter than T$_{1,bulk}$ (typically a few ms, even in the presence of other bulk paramagnetic impurities). We attribute this difference to unpaired paramagnetic spins on the ND surface~\cite{NDsurface1,NDsurface2,SurfaceVib1,SurfaceVib2}.  
The magnetic noise induced by these surface spins adds a new depolarization channel and yields a decrease in T$_1$ according to Eq.~(\ref{eq:fermigoldenrule}).
Based on this model, we can further deduce the surface spin density to be about 1 nm$^{-2}$~\cite{SI}. The T$_{1}$ relaxometry can then provide information in estimating surface spin densities of ND samples. 

We first show the strong T$_{1}$ quenching effect induced by Gd molecules in the solution. According to Eq.~(\ref{eq:fermigoldenrule}), the presence of Gd molecules will induce a strong magnetic noise in addition to that arising from the surface spins, and significantly increase the sensor's spin relaxation rate $1/T_1$. 
We prepared solutions of NDs  and varying densities of Gd-DOTAs,  and measured the corresponding T$_1$~of the NV centers. 
Higher densities of Gd-DOTA molecules lead to stronger magnetic noise thus larger T$_{1}$ quenching ratio. 
The experimental results quantitatively match our theoretical predictions if we assume a higher Gd density than the average density in solution (Fig.~\ref{Fig2}). 
This is likely due the tendency of Gd molecules to accumulate close to the bottom of the microfluidic channel. Since we measured NV spins at this location, this accumulation yields higher Gd density close to detection spots than the average concentration in the solution. Then, as the sensors only detect the local rather than global averaged environment, they see an effective higher Gd molecule density.

We next show the capability of detecting the RBMs of Gd-DOTA molecules with our sensors. In particular, we are interested in demonstrating the ability to distinguish variations in the magnetic particle fluctuation rates due to changes in the solution viscosity.
We dissolved a fixed density of Gd-DOTA molecules in solutions with varying concentrations of water and acetone. 
Varying the ratio between water and acetone in the binary solution  changes the local viscosity felt by Gd-DOTA molecules. 
We perform T$_1$ relaxation measurements by loading the sample in a microfluidic channel to mimic biological environment, while preventing the binary solution from quickly evaporating.
We observed a clear difference (4.34 $\sigma$) in the measured T$_1$ times for pure water   with respect to (nearly) pure acetone (Fig.~\ref{Fig3}~(a)). To understand more quantitatively these results, we develop an analytical model of the RBM rate in our experiments.

The RBM of a molecule is influenced by its local viscosity. For a molecule with hydrodynamic radius $a$ in a solvent of viscosity $\eta$ and molecular radius $a_{s}$, the RBM rate is expressed by the Stokes-Einstein equation~\cite{Rast2000,MerbachBook2013}:
\begin{eqnarray}
\label{Rrot}
R_{rot}=\frac{k_{B}T}{8\pi a^{3}\eta f_{r}}, \qquad f_{r}=\left(\frac{6a_{s}}{a}+\frac{1+\frac{3a_{s}}{a+2a_{s}}}{(1+\frac{2a_{s}}{a})^{3}} \right)^{-1}
\end{eqnarray}
Here $f_{r}$ is the microviscosity factor that takes into account the discrete nature of solvent molecules~\cite{Rast2000,Fries1995}. As we can see, a lower viscosity corresponds to faster motions, with a  RBM rate $R_{rot}$  as large as $14$~GHz in our experiment. 
We plot in in Fig.~\ref{Fig3} (b) the dynamical viscosity $\eta$ and the expected RBM rate of Gd-DOTA. The RBM rate of Gd-DOTA molecules  varies between around $2$ to $14$~GHz in our water-acetone binary solution: This large range is attributed to the contrast in viscosities as well as different molecule sizes of water and acetone. 
Since $R_{Gd,tot}\gg \omega_0$ due to the large $R_{Gd,dip}$ contribution, we expect longer relaxation time of the NV sensors for higher RBM rates, as seen in Fig.~\ref{Fig3}~(a).

We further verified that our assumption that the  microviscosity factor is essential to describe the RBM of the (nanoscale) Gd-DOTA molecules is correct. We compared the estimated RBM rate using Eq.~(\ref{Rrot}), either with and without the microviscosity factor (Fig.~\ref{Fig3} (c,d)), and found the predicted RBM rate with the microviscosity factor agrees better with experimental results (Fig.~\ref{Fig3} (a)). This matches our expectation, since the discrete nature of solvent molecules should come into play when sub-nanometer sized Gd-DOTA molecules are concerned.
Finally, from  the experimental data and considering a $10 \%$ variation in Gd molecule density, we estimate that the RBM rate of Gd molecules in nearly pure acetone is $20.2\pm10.6$~GHz, consistent with the $14.2$~GHz calculated from Eq.~\ref{Rrot} at room temperature. 

Since our measurements were performed with an ensemble of NDs, one concern was the inhomogeneous density distribution of Gd molecules when one changes solutions. To investigate this issue, we compared the distribution of relaxation times of spatially separated sensing spots for the water and nearly pure acetone cases. As one can see in Fig.~\ref{Fig4}~(a), the two distributions spread broadly, which might be a result of the spatial inhomogeneity of Gd molecule density or spatially varying local charge environments (see next section). Nevertheless,  the two distributions are clearly distinct form each other, as shown in Fig.~\ref{Fig4}~(b), with Gaussian fittings. This further demonstrates that our relaxation measurements can distinguish different RBM rates influenced by local viscosities. 

\section{Discussions}
 
To quantify the performance of our proposed method,  we can estimate the minimal detectable value of the total magnetic noise fluctuation rate $\delta R^{min}_{Gd}\sqrt{T}$ per unit time, for a single NV center located at the center of a single ND. We find
  \begin{eqnarray}\label{eq:motionalrate}
  \delta R^{min}_{Gd}\sqrt{T} \approx  \frac1{C\sqrt{\mathscr{D}T_D}}\sqrt{\frac{2e R_{Gd,tot}}{3 \gamma_{e}^{2}B_{\perp,Gd}^2}}\frac{(R_{Gd,tot}^{2}+\omega_{0}^{2})^{3/2}}{ |R_{Gd,tot}^{2}-\omega_{0}^{2}|},
 \end{eqnarray}
where $T_D$ is the detection window, $\mathscr{D}$ the photon counting rate, and $C$ the contrast.
In our experiments we have a detection window $T_{D} = 500$ ns. We assume a signal contrast $C=0.2$ and a photon counting rate $\mathscr{D}=1 \times 10^{5}$ counts/s for a single NV. At optimized Gd density (corresponding to total Gd fluctuation rate  $R_{Gd,tot}\approx 60.2$ GHz), we get a sensitivity of $ \delta R^{min}_{Gd}$=6.9 (9.6) GHz for a single ND with diameter $d_{0}$=20 (25) nm in a $T=10 $ s data acquisition time, as presented in Fig.~\ref{Fig4} (c). Note that here the calculation only takes the photon shot-noise into consideration, as it is the main source of detection noise.

To extend our method to single  NV measurements, which can yield superior spatial resolutions, we need to have well-characterized NV centers with spatially homogeneous distributed magnetic molecules around them. In addition,  
the charge state conversion between NV$^{-}$ and undesired NV$^{0}$ states should be well characterized. 
 Indeed, the charge conversion rate might vary both due to different chemical solutions and across NV centers, thus masking the real relaxation differences. As we averaged over many NDs, this was not a main concern here. We point out that this issue can be addressed by characterizing or mitigating the charge instabilities beforehand with the use, for example, of additional lasers to control the photoionization process, or microwave control to better select the NV$^-$ dynamics~\cite{PhysRevLett.122.076101,JorgNJP2013,EnglundPNAS2016}.

We note that the presence of carboxyl termination group on the surface of NDs might naturally compete and interact with the DOTA chelators.  Careful surface treatment should be performed to reduce this effect and increase the colloidal stability of NDs to make the technique more suitable for practical applications.  
For example, fluorination of NDs can suppress detrimental hydrophobic interactions and make NDs’ aqueous solution colloidally stable~\cite{NDsurfaceFluorination,HemmerNDreview2018}.  
The method we proposed here is capable of determining the RBM rate (thus viscosity and particle hydrodynamic size) and, subject to careful evaluations of all assumptions in our model~\cite{SI}, we can turn our qualitatively measurements into quantitative ones, while improving the sensitivity. 
Future studies that independently determine particle accumulation in the microfluidic channel could also help improve the performance of our sensor.

We have shown that, in the context of quantum sensing, the T$_{1}$ relaxometry provides a versatile tool to detect the RBM rate changes in response to variations of local environments. At the same time,  we point out that the relation between sensor relaxation and rate variation can be employed in MRI techniques, where the fluctuation rates of contrast agents (usually magnetic molecules such as Gd(III) chelators) are modulated to increase their relaxivity and improve their performance~\cite{MerbachBook2013}.

\section{Conclusions}

In conclusion, we proposed a novel method to detect the rotational Brownian motion of nanometer-sized magnetic particles, and experimentally demonstrated the protocol using NV centers in nanodiamonds. The technique is capable of detecting fast rotation (GHz-rate) of magnetic molecules with size down to sub-nanometer scale. The experiment is fully optical, requiring no microwave control or any external magnetic field. Our quantum sensing technique provides a new way of extracting local viscous information with high spatial resolution,  and could also contribute to the design of contrast agents for MRI.

\textbf{\textit{Acknowledgements }}

We thank Tingtao Zhou and Yixiang Liu for fruitful discussions, Liyuan Zhang and Yinan Shen for help with DLS measurement and Kurt Broderick for assistance with the fabrication of microfluidic channels. 
This work was supported in part by the U.S. Army Research
Office through Grants No. W911NF-11-1-0400 and No. W911NF-15-1-0548. 
DL acknowledges financial support from the German Research Foundation (DFG) with a postdoctoral fellowship.

\medskip
\bibliography{GdRBM_arXiv,Biblio} 
\bibliographystyle{unsrt}

\newpage


\section*{Supplementary material}

# Supporting material for "All-optical quantum sensing of rotational Brownian motion of magnetic molecules"

Changhao Li[1,2], Mo Chen[1,3], Dominika Lyzwa[1], and Paola Cappellaro [*,1,2]

[1]*Research Laboratory of Electronics, Massachusetts Institute of Technology, Cambridge, MA 02139*
[2]*Department of Nuclear Science and Engineering, Massachusetts Institute of Technology, Cambridge, MA 02139*
[3]*Department of Mechanical Engineering, Massachusetts Institute of Technology, Cambridge, MA 02139*

# Contents

---
[*]pcappell@mit.edu

# 1 Experiment

## 1.1 Sample preparation

We used nanodiamonds (NDs) from Adamasnano that were milled from high pressure high temperature (HPHT) diamond particles which had substitutional nitrogen concentration around 100-120 ppm. The diamond particles were irradiated with 2-3 MeV electron with doses ranging from $5 \times 10^{18}$ to $5 \times 10^{19}$ $e/cm^2$ to create vacancies. Then a 2-hour annealing process at 850 °C was performed to bring vacancies to substitutional nitrogen atoms in diamond lattice to form nitrogen-vacancy (NV) centers. The NDs were supposed to have 1-4 NV emitters inside each NDs on average. They were carboxylic acid groups enriched by oxidation in nitric and sulfuric acid hence they were terminated with carboxyl groups.

The nanodiamonds we used have an average size from 20 to 25 nm. We measured the size distribution of our ND sample in water solution at 25 °C with a dynamical light scatter (DLS, Zetasizer nano), as shown in Fig. S1. Single NDs were further characterized with a transmission electron microscopy (TEM) at 200 keV (JEOL 2010) (Fig. S2).

Our Gd chelator molecules are Gd-DOTA (Gadolinium (III) 1,4,7,10-Tetraazacyclododecane-1,4,7,10-tetraacetate) purchased from Macrocyclics. We dissolved the molecules in water and sonicated to mix them well. The stability of macrocyclic ligands such as DOTA here is usually higher than that of a complex formed with an open chain ligand owing to the macrocyclic effect [1]. The ultra-low concentration of free Gd ions (ions without chelation) in the solution under equilibrium greatly reduces their toxicity as contrast agents in practical applications such as MRI. We note that the stability of the DOTA complexes might be degraded due to its competition with the carboxyl groups on the surface of nanodiamonds.

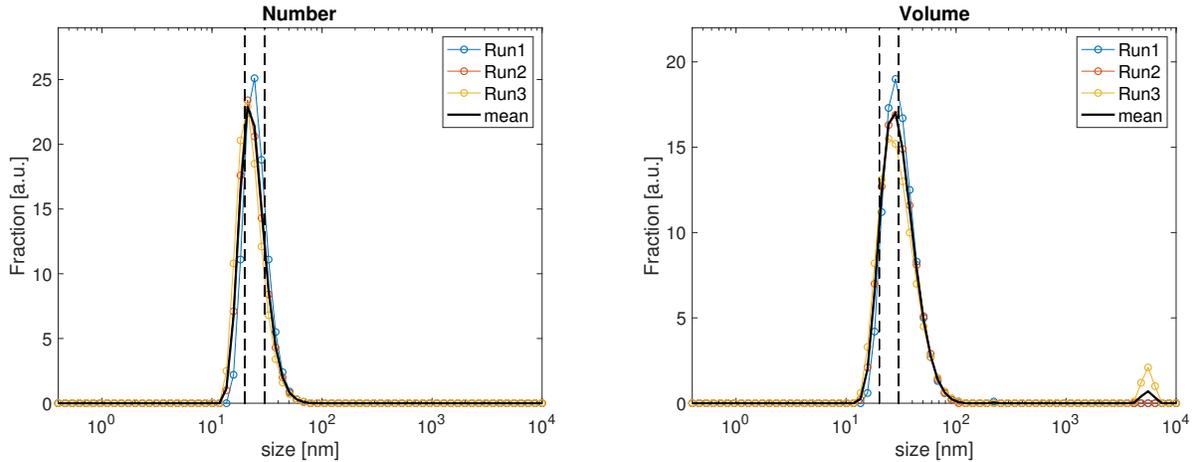

Figure S1: Size distribution of nanodiamonds. We use dynamical light scatter to measure the number (left) and volume (right) distributions and they show peaks between 20 nm and 30 nm (dashed vertical lines). Note that the volume distribution might overestimate the size of the particles.

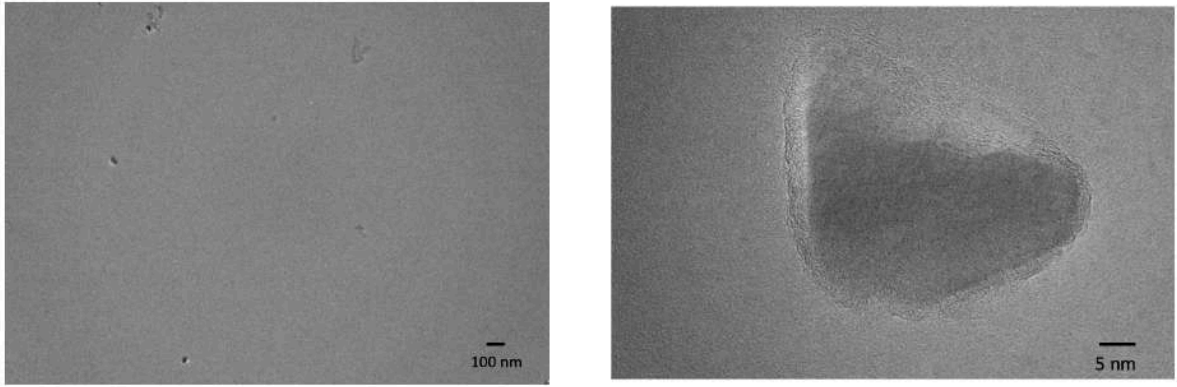

Figure S2: TEM characterization of nanodiamonds. ND solution was dripped in a copper grid with a thin carbon film and dried before they were ready for TEM measurements.

## 1.2 Experimental setup

Our optical measurements were performed in a homebuilt confocal microscope with a NA=0.95 air objective. An acousto-optic modulator (Isomet) swtiches the 532 nm green excitation on (off) and the red fluorescence is collected by a single photon detector (PerkinElmer). When measuring the rotational rates of magnetic molecules in binary solutions, we loaded the sample into a microfluidic channel made of transparent polydimethylsiloxane (PDMS, from Sylgard 184 silicone elastomer kit, Dow Corning) to mimic biological environment. The PDMS channel was bonded to a cover glass after oxygen plasma treatment. Plastic tube and syringe were used as inlet and outlet systems to load the solution (Fig. S3). Note that the channel can significantly slow down the evaporation of acetone solvents, an undesired process which can change the density of Gd molecules. The channel we used in the above experiments had a width of around 150 $\mu$m which were confirmed by confocal images (Fig. S3) and the height was around 40 $\mu$m.

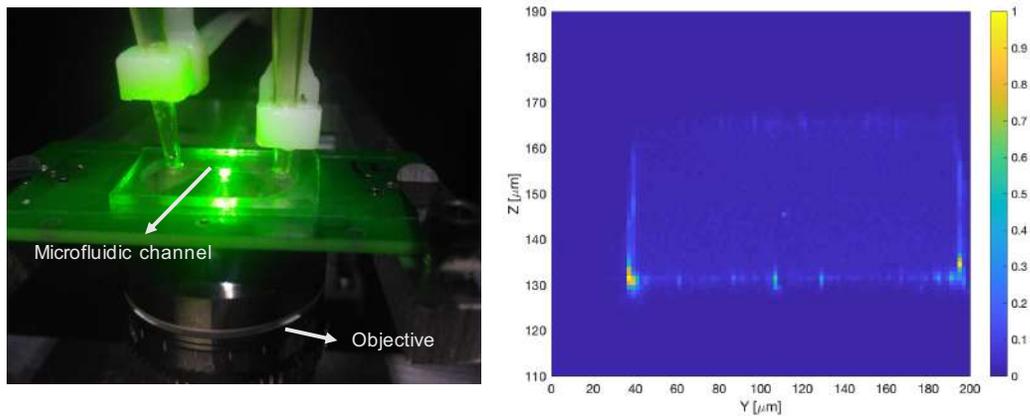

Figure S3: Real (left) and confocal cross-sectioned (right) image of microfluidic channel. The bright area near the edges of the channel in the confocal image is due to the accumulation of nanodiamonds.

3

## 1.3 Viscosity measurement

We performed viscosity measurements of the water-acetone binary solutions where the acetone we used (Macron Fine Chemicals, 2440-16) has weight concentration over 99 %. The viscosity of the mixtures was measured with a size 0 Ubbelohde viscometer (Cannon Instrument) at 20.4 °C. We take the known kinetic viscosity of pure acetone as a reference to calculate the capillary constant of the viscometer. The measured data, is compared to reference [2] and shown below (Fig. S4). Note that the viscosity peaks near water mole fraction 0.8 and this should be due to the formation of water-acetone micelles, which also induces a positive viscosity derivation from the expected linear curve [2]. The micelle, where the polar heads of acetone molecules form the surfaces, is surrounded by water molecules, thus forming a cluster consisting of the micelle and water layers.

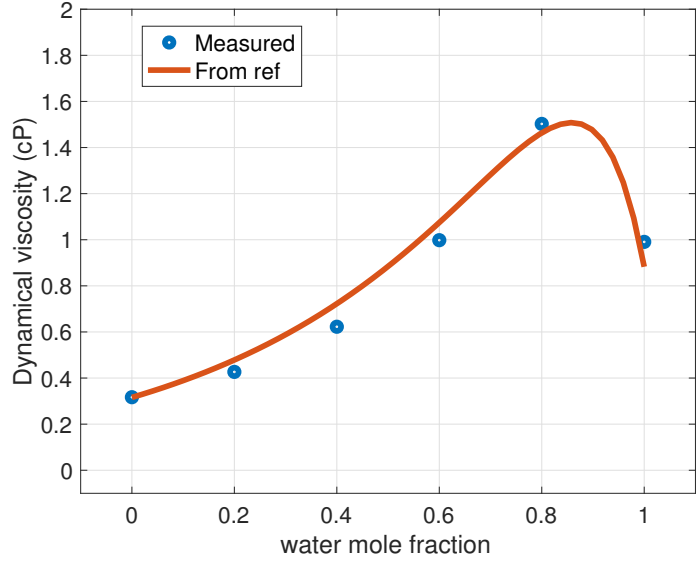

Figure S4: Viscosity of water-acetone solutions. The red curve represents the data from the reference [2] which agrees well with our measurements (blue circles).

## 2 Theoretical derivation of the relaxation time

In this section, we give the detailed derivation of our model for calculating the NV center's relaxation time. While some of the results below have been derived in reference [3], we briefly repeat them here for the convenience of readers.

### 2.1 $T_1$ decay signal

With the NV spin initially polarized by green laser and in the absence of any external magnetic field, the probability of finding NV state in $|m_s = 0\rangle$ or $|m_s = \pm 1\rangle$ is given by:

$$P_0(\tau) = \tfrac{1}{3} + (P_0(0) - \tfrac{1}{3})e^{-\tau/T_1} \tag{S1}$$
$$P_{\pm 1}(\tau) = \tfrac{1}{2}(1 - P_0(\tau)). \tag{S2}$$

4

At thermal equilibrium, the three NV ground states will have equal occupation probability 1/3. The measured relaxation signal is given by:

$$\mathscr{S}(\tau) = F_0 P_0(\tau) + F_1(P_{-1}(\tau) + P_{+1}(\tau)) \quad (S3)$$

where $F_0$ and $F_1$ are the fluorescence counts for $|m_s = 0\rangle$ and $|m_s = \pm 1\rangle$ respectively. Due to the non-radiative transition process mentioned in the main text, we have $F_0 > F_1$. Then the signal can be written as:

$$\mathscr{S}(\tau) = \mathscr{S}_\infty (1 + C e^{-\tau/T_1}) \quad (S4)$$

with $\mathscr{S}_\infty = \frac{F_0 + 2F_1}{3}$ the signal counts at thermal equilibrium and the contrast $C$ given by:

$$C = (3P_0(0) - 1)\frac{F_0 - F_1}{F_0 + 2F_1} \quad (S5)$$

which is related to the initial polarization fidelity $P_0(0)$ and fluorescence intensity difference between different spin states $F_0 - F_1$. We note that the relaxation from the non-radiative metastable state can result in a further modulation of the signal [3], but the relevant time scale (typically hundreds of nanoseconds) is considerable smaller than the timescale $T_1$ which is of interest in our experiments.

## 2.2 $T_1$ decay formalism

From Fermi's golden rule, one can estimate the relaxation rate [3]:

$$\frac{1}{T_1} = \frac{1}{T_{1,bulk}} + \sum_k 3\gamma_k^2 B_{\perp,k}^2 \frac{\tau_{c,k}}{1 + \omega_0^2 \tau_{c,k}^2} \quad (S6)$$

where $\omega_0/2\pi = 2.87$ GHz is the NV zero-field splitting (we neglect the zero-field splitting of Gd which is much smaller than $\omega_0$), $T_{1,bulk}$ is the relaxation time for NV centers in bulk diamond at room temperature and we take it to be 1 ms. Here the sum over $k$ represents the sum of different noise sources, for example, surface spins and Gd spins.

For the spins on the diamond surface, the correlation time $\tau_{c,Surf}$ is mainly from the intra-bath dipole-dipole interactions between each other, and their vibrational spin relaxation can usually be neglected according to EPR studies [4, 5]. That is,

$$\tau_{c,Surf} \approx 1/R_{dip,Surf}. \quad (S7)$$

For the Gd spins, apart from the vibrational and dipolar rates, there are additional fluctuation sources induced by random motions of Gd molecules in the solution. As we will show in the sections below,

$$\tau_{c,Gd} = 1/R_{Gd,tot} = 1/(R_{dip,Gd} + R_{vib} + R_{trans} + R_{rot}). \quad (S8)$$

We will investigate each terms in Eq. S6 and Eq. S8 in the following. In spite of the assumptions in the following derivations, we expect that our result captures the essence of the physical process and provides a correct order of magnitude estimation.

5

## 2.3 Transverse magnetic noise

The transverse magnetic noise from the environment induces the longitudinal relaxation of central spins (NV centers). We consider a single NV center located at the center of a ND of diameter $d_0$. Let's first calculate the transverse field $B_\perp^2$ by summing up contributions from all bath spins:

$$B_\perp^2 = \sum_i B_{\perp,i}^2 = \sum_i Tr(\rho_{th} B_{x,i}^2 + \rho_{th} B_{y,i}^2) \tag{S9}$$

where $\rho_{th}$ is the fully mixed thermal state (identity matrix) for a spin with spin number S. The noise strength $\boldsymbol{B}_i$ from spin $\boldsymbol{S}_i$ located at position $\boldsymbol{r}_i$ is given by the dipolar force felt by the NV spin at the center:

$$\boldsymbol{B}_i = \frac{\mu_0 \gamma_i \hbar}{4\pi r_i^3} \left[ \boldsymbol{S}_i - 3\left(\boldsymbol{S}_i \cdot \frac{\boldsymbol{r}_i}{r_i}\right) \frac{\boldsymbol{r}_i}{r_i} \right] \tag{S10}$$

Then one can evaluate the transverse field:

$$B_\perp^2 = \sum_i B_{\perp,i}^2 = \sum_i \left(\frac{\mu_0 \gamma_i \hbar}{4\pi}\right)^2 C_s \frac{2+3\sin^2\theta_i}{r_i^6} \tag{S11}$$

where $C_s = \frac{S(S+1)}{3}$ is determined by the spin types, $\gamma_i$ is the gyromagnetic ratio of spin $i$ and $\theta_i$ is the angle between the NV quantization axis and the relative direction of spin $i$ with respect to the central NV spin. In the pure ND case, we only consider the surface spin (assuming spin 1/2) with density $\sigma$, then we can get:

$$\begin{aligned} B_{\perp,Surf}^2 &= \left(\frac{\mu_0 \gamma_e \hbar}{4\pi}\right)^2 C_{s,Surf} \int_0^{2\pi} d\phi \int_0^{\pi} (2+3\sin^2\theta)\sin\theta d\theta \lim_{\Delta r \to 0} \frac{\int_{d_0/2}^{d_0/2+\Delta r} \frac{r^2}{r^6} dr}{\Delta r / \sigma} \\ &= \left(\frac{\mu_0 \gamma_e \hbar}{\pi}\right)^2 C_{s,Surf} \pi \frac{\sigma}{(d_0/2)^4} \end{aligned} \tag{S12}$$

where $d_0$ is the diameter of ND and $C_{s,Surf} = 0.25$, assuming $S = 1/2$ for surface spins.

Similarly, for the noise strength induced by Gd molecules with density $n$:

$$\begin{aligned} B_{\perp,Gd}^2 &= \left(\frac{\mu_0 \gamma_{Gd} \hbar}{4\pi}\right)^2 C_{s,Gd} \int_0^{2\pi} d\phi \int_0^{\pi} (2+3\sin^2\theta)\sin\theta d\theta \frac{\int_{d_0/2}^{+\infty} \frac{r^2}{r^6} dr}{1/n} \\ &= \left(\frac{\mu_0 \gamma_{Gd} \hbar}{\pi}\right)^2 C_{s,Gd} \pi \frac{1}{(d_0/2)^4} \frac{n d_0}{6} \end{aligned} \tag{S13}$$

with $C_{s,Gd} = 10.5$ (note that Gd has a spin 7/2). For a single NV center at the center of a ND with diameter 20 (25) nm, we estimate the transverse noise to be $B_{\perp,Gd} \approx 778$ (557) $\mu$T $\times n^{1/2}$. We can also get the ratio of the transverse field from transverse and surface spins.

$$\frac{B_{\perp,Gd}^2}{B_{\perp,Surf}^2} \approx \frac{n d_0}{6\sigma} \frac{C_{s,Gd}}{C_{s,Surf}} \left(\frac{\gamma_{Gd}}{\gamma_e}\right)^2 \tag{S14}$$

Given the fact that the g-factor for Gd complex is approximately that of the surface electron spin's [6], here we take $\gamma_{Gd} = \gamma_e$.

## 2.4 Dipolar interactions between nearby magnetic spins

Dipolar interactions between magnetic spins is one of the main contributions to the magnetic noise fluctuations, shortening their correlation time. We calculate this fluctuation rate by computing the second moment of the dipolar interaction. For a system at infinite temperature, $M_2 = \text{Tr}(H_{dip}^2)$ [7]. $H_{dip}$ is the sum over the dipolar interactions $H_{ij}$ between spin $i$ and $j$ at distance $r_{ij}$:

$$H_{ij} = \frac{\mu_0 \gamma_e^2}{4\pi r_{ij}^3} \left[ \boldsymbol{S}_i \cdot \boldsymbol{S}_j - 3 \left( \boldsymbol{S}_i \cdot \frac{\boldsymbol{r}_{ij}}{r_{ij}} \right) \left( \boldsymbol{S}_j \cdot \frac{\boldsymbol{r}_{ij}}{r_{ij}} \right) \right] \quad (S15)$$

The dipolar fluctuation rate is then:

$$\hbar R_{dip} = \sqrt{\sum_{i \neq j} \langle H_{ij}^2 \rangle} = \frac{\mu_0 \gamma_e^2 \sqrt{6} C_s}{4\pi} \left( \sum_{i \neq j} \frac{1}{r_{ij}^6} \right)^{1/2} \quad (S16)$$

We first consider the surface spins. By assuming a surface density $\sigma$, we can rewrite the sum as an integral. The distance between spins on the diamond surface is $r_{ij}^2 = 2r^2(1 + \cos\theta_{ij})$, where $r = d_0/2$ is the ND radius and $\theta_{ij}$ is the angle between the spins with respect to the NV center. Then we can integrate over the surface: $\sum_{i \neq j} \frac{1}{r_{ij}^6} = \sigma \int_{\theta_{min}}^{\pi/2} 2\pi r^2 \frac{\sin\theta d\theta}{8r^6(1+\cos\theta)^3} = \sigma\pi(1/2r_{min}^4 - 1/8r^4)$, where $r_{min} \approx 0.15$ nm is the minimum allowed distance between surface spins, estimated by the nearest neighbor lattice distance in diamond. Since this is much smaller than the ND radius, we have $\sum_{i \neq j} \frac{1}{r_{ij}^6} \approx \frac{\pi}{2} \frac{\sigma}{r_{min}^4}$. Then,

$$\hbar R_{dip,Surf} = \sqrt{\sum_{i \neq j} \langle H_{ij}^2 \rangle} = \frac{\mu_0 \gamma_e^2 \hbar^2 \sqrt{6} C_{s,Surf}}{4\pi} \left(\frac{\pi}{2}\right)^{1/2} \frac{\sigma^{1/2}}{r_{min}^2} \Rightarrow R_{dip,Surf} \approx 11 \quad \text{ns}^{-1}\text{nm} \times \sigma^{1/2}. \quad (S17)$$

For Gd spins with density $n$, similarly we have $\sum_{i \neq j} \frac{1}{r_{ij}^6} \approx n \int_{r_{min}}^{+\infty} \frac{4\pi r^2}{r^6} dr \approx \frac{4\pi n}{3 r_{min}'^3}$ where, again we take $r_{min}' \approx r_{min}$ in our model. This estimation could be further improved by considering the actual size and the anisotropy of Gd molecules. We then reach the dipolar interaction rate:

$$\hbar R_{dip,Gd} = \sqrt{\sum_{i \neq j} \langle H_{ij}^2 \rangle} = \frac{\mu_0 \gamma_e^2 \hbar^2 \sqrt{6} C_{s,Gd}}{4\pi} \left(\frac{4\pi n}{3 r_{min}'^3}\right)^{1/2} \Rightarrow R_{dip,Gd} \approx 296 \quad \text{ns}^{-1}\text{nm}^{3/2} \times n^{1/2}. \quad (S18)$$

## 2.5 Estimation of surface spin density

In the absence of Gd molecules, the relaxation time of NV centers in NDs is still significantly shorter than the ones in bulk diamond which is typically between 1 ms and 10 ms at room temperature. Note that the bulk NV $T_1$ time is limited by phonon processes or other defects in diamond such as substitutional nitrogen atoms (P1 centers) [8, 9], and here we take a bulk relaxation time of 1 ms.

The measured short relaxation time for NV centers in NDs can be ascribed to the bath of paramagnetic surface spins, for example, dangling bonds with unpaired electrons with spin 1/2. From the relaxation time dependence on surface spin density $\sigma$ (Fig. S5) derived above, we estimated $\sigma \approx 1$ nm$^{-2}$ in our ND samples.

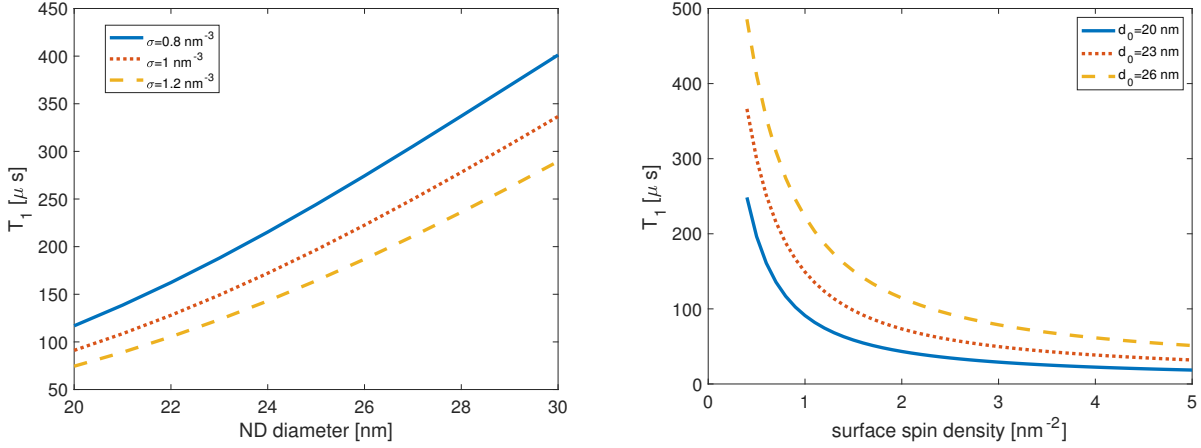

Figure S5: Estimation of NDs' surface spin density. Here we assume no Gd molecules were added. *Left:* $T_1$ as a function of ND diameter for fixed surface spin densities. *Right:* $T_1$ as a function of surface spin density for fixed ND diameters. From our measured relaxation time $T_1 \approx 133\mu s$, we estimated $\sigma \approx 1 \text{nm}^{-2}$ in our samples.

## 3 Fluctuation rate of Gd molecules

Here we consider Gd-DOTA complexes which are the samples we used in the experiments and are commonly used as contrast agents in MRI. Recall that the spectral density $S_{Gd}(\omega)$ has a Lorentzian lineshape with its width determined by the total fluctuation rate $R_{Gd,tot}$:

$$S_{Gd}(\omega) \propto \frac{R_{Gd,tot}}{R_{Gd,tot}^2 + \omega_0^2} \tag{S19}$$

Here $R_{Gd,tot} = R_{dip,Gd} + R_{vib} + R_{trans} + R_{rot}$. We have shown that the first term is given by Eq. S18.

The vibrational term $R_{vib}$ originates from the electronic spin relaxation of Gd itself. The relaxation rate is ascribed to transient zero-field splitting (ZFS) induced by solvent collisions or molecular vibrations. The value of $R_{vib}$ at zero external field is then determined by the correlation time of the fluctuation of the transient ZFS $\tau_v$, and trace value of ZFS tensor $\Delta$: $R_{vib} = 1/\tau_{s0} = 12\Delta^2\tau_v$ [1, 10]. Then for Gd-DOTA at 298 K, we have $R_{vib} = 1/473\,\text{ps}^{-1} \approx 2.1$ GHz [10].

The last two terms, $R_{trans}$ and $R_{rot}$, are due to the Brownian motion of the molecules. The fluctuation rate of the magnetic noise induce by Gd molecules' translational diffusion is:

$$R_{trans} = D_{\text{diff}} \left(\frac{3}{4r}\right)^2 = 1.69 \times \text{GHz}\,\text{nm}^2 \times \frac{1}{r^2} \tag{S20}$$

if one assumes the typical diffusion coefficient $D_{\text{diff}} = 3 \times 10^{-9} m^2 s^{-1}$ at room temperature. Here $r$ is the sensor-sample distance and $r \geq \frac{d_0}{2} \approx 12$ nm for our diamond sample. This yields a translational motional rate less than 12 MHz, which is negligible compared to other terms. The intuition here is that the larger the NV-Gd distance becomes, the less sensitive to the Gd molecule's diffusion the NV center is.

8

The rotational fluctuation rate of Gd molecules can be derived from the Stokes-Einstein equation [11]:

$$R_{rot} = \frac{k_B T}{8\pi a^3 \eta f_r} \tag{S21}$$

where $a$ is the radius of the complex and $\eta$ the solution viscosity. Note for water the viscosity is 0.89 cP, while it is 0.32 cP for acetone at room temperature. As verified in our experiments, in micro-heterogeneous solutions we need to insert a microviscosity factor $f_r$ [11, 12]:

$$f_r = \left( \frac{6a_s}{a} + \frac{1 + \frac{3a_s}{a+2a_s}}{(1+\frac{2a_s}{a})^3} \right)^{-1}, \tag{S22}$$

where $a_s$ is the solution molecule radius (for water it is $a_s = 0.14$ nm). This factor takes the discrete nature of the solvent and solute molecules into account. Note it will increase as a function of the ratio $a/a_s$, when the solute size increases and approaches unity for large solute particles.

The rotational fluctuation rate is roughly inversely proportional to the hydrodynamic volume of the complex. For $[Gd(H_2O)_8]^{3+}$ ($a = 0.39$ nm), the estimated fluctuation rate in water is 7.65 GHz at room temprature [11]. With the ratio of volumes of $[Gd(DOTA)(H_2O)]^-$ over $[Gd(H_2O)_8]^{3+}$ estimated from corresponding Connolly surfaces to be 2.3 [13, 14], the former has $a = 0.51$ nm and then $f_r = 0.486$. Considering the factor $f_r$, the rotational rate of the Gd-DOTA should be around 2.75 GHz, while independent measurements in the literature [14] reported a value around 2 GHz at room temperature, in line with our calculation here.

In Fig. S6 we show the different fluctuation sources of Gd-DOTA molecules discussed above as a function of molecule density $n$ in pure water. At low density, the rotational rate term can be significant.

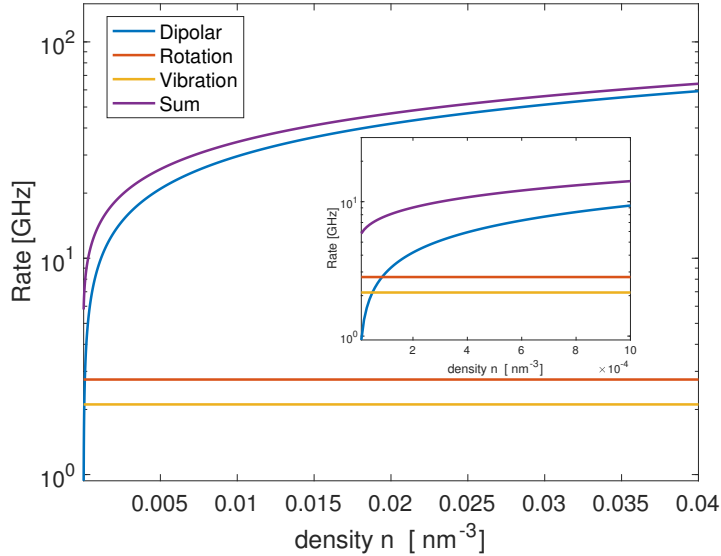

Figure S6: Fluctuation rates of Gd molecules from different mechanisms, as a function of Gd density, at room temperature. Figure inset shows the rates at low densities.

Since $R_{Gd,tot}$ is usually larger than $\omega_0$ (here $\omega_0/2\pi = 2.87$ GHz) due to the large dipolar term, as discussed in main text, we expect that a further increase of fluctuation rate will increase the $T_1$ thus decreasing the relaxivity. In practice applications such as MRI, at a relative low magnetic field, one

can attach Gd complex with polymers or protein to increase the molecule size, thus slowing down the rapid tumbling and enabling higher relaxivity [1].

## 4 Effect of laser heating

The laser used to polarize and readout the NVs increases the local temperature of the detected NDs as well as Gd molecules. The NDs are stable against temperature changes and the NV bare $T_1$ does not change too much [9]. The complex molecule we used, Gd-DOTA, possesses a high thermal stability under ambient conditions as well. Gd-DOTA molecules are also kinetically inert at a relative high temperature so they do not dissociate to release free Gd ions. Hence the structure of the molecules will not change under laser heating.

The temperature fluctuation can only tune the zero-field level splitting of NV centers slightly. The transition frequency $\omega_0$ has a temperature dependence $d\omega_0/dT = -2\pi \times$ 77 kHz K$^{-1}$ due to thermally induced lattice strains, hence a temperature change of 10 K corresponds to energy level change less than 1 MHz which is much less than Gd noise spectrum width.

Meanwhile, the rate of rotational Brownian motion and intrinsic vibration of Gd molecules have a temperature dependence. In Fig. S7 we show the rates for pure water and pure acetone case. While the rotational rate increases with higher temperature, the vibrational rate of Gd molecules shows the opposite trend and the sum of the two only differ by less than 1 GHz in the shown temperature range.

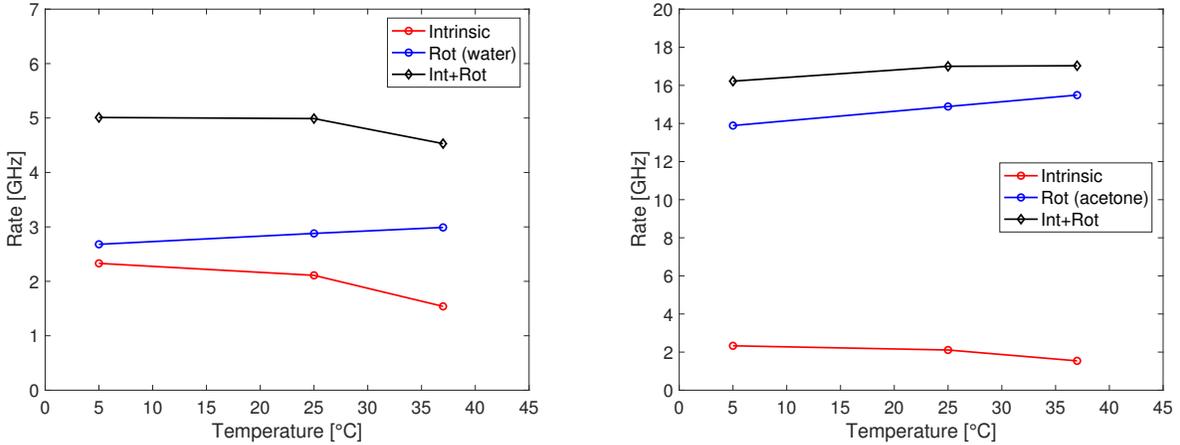

Figure S7: Fluctuation rates of Gd molecules for pure water (left) and (nearly) pure acetone (right). The rotational Brownian motion rates are calculated from Eq. S21 and the intrinsic vibrational rates are fitted data from NMRD experiments in reference [10].

## 5 Sensitivity estimation

In this section we derive the sensitivity of our protocol to changes in the total fluctuation rate of Gd complexes $R_{Gd,tot}$ and then to the solution's viscosity $\eta$, assuming a single NV center at the center of a nanodiamond.

10

## 5.1 Sensitivity to fluctuation rate

We consider performing $N$ measurements on a single NV center for a total time $T$, each measurement detecting the spin relaxation after a time $\tau$ (we neglect dead times associated with polarization and detection of the NV centers, as these are small with respect to $\tau$). The signal, that is, the total number of detected photons, is given by:

$$N(\Gamma_1) = \mathscr{D} T_D \frac{T}{\tau} \left(1 + C e^{-\Gamma_1 \tau}\right), \tag{S23}$$

where $T_D = 500$ ns is the detection window, $\mathscr{D}$ the photon detection rate and $C \approx 0.2$ is the signal contrast for a typical single NV measurement. Here $\Gamma_1 = 1/T_1$ is the relaxation rate as given in Eq. S6:

$$\Gamma_1 = \Gamma_1^{const} + 3\gamma_e^2 B_{\perp,Gd}^2 \frac{R_{Gd,tot}}{R_{Gd,tot}^2 + \omega_0^2}, \tag{S24}$$

where $\Gamma_1^{const}$ is the (constant) relaxation rate contributions from bulk and surface spins. A variation in the Gd fluctuation rate $\delta R_{Gd}$ will induce a change in the relaxation $\Gamma_1 \to \Gamma_1'$, thus changing the number of detected photons. If the variation is small, we can consider just the first order:

$$\Gamma_1' = \Gamma_1 + \delta\Gamma_1 = \Gamma_1 + 3\gamma_e^2 B_{\perp,Gd}^2 \frac{|\omega_0^2 - R_{Gd,tot}^2|}{(R_{Gd,tot}^2 + \omega_0^2)^2} \delta R_{Gd} \tag{S25}$$

The associate signal variation is $\delta N = N(\Gamma_1) - N(\Gamma_1')$,

$$\delta N = \mathscr{D} T_D \frac{T}{\tau} C e^{-\Gamma_1 \tau} (1 - e^{-\delta\Gamma_1 \tau}) \approx \mathscr{D} T_D T C e^{-\Gamma_1 \tau} \delta\Gamma_1 \tag{S26}$$

Since the photon shot noise dominates the experiments, we can take the noise to be $N_{noise} = \sqrt{N(\Gamma_1)} \approx \sqrt{\mathscr{D} T_D T/\tau}$, where we neglected contributions from $C$ as the contrast is typically small, $C \ll 1$. We thus obtain the signal-to-noise ratio (SNR):

$$\text{SNR} = \frac{\delta N}{\delta N_{noise}} = \sqrt{\mathscr{D} T_D T \tau}\, C \delta\Gamma_1 e^{-\Gamma_1 \tau} \tag{S27}$$

The SNR is maximum at $\tau = T_1/2$,

$$\text{SNR}_{max} = \frac{\delta\Gamma_1}{\sqrt{\Gamma_1}} C \sqrt{\frac{\mathscr{D} T_D T}{2e}} \tag{S28}$$

Then, the minimum fluctuation rate change $\delta R_{Gd}^{min}$ that can be detected is obtained for a SNR of one. We thus obtain the sensitivity per unit time:

$$\delta R_{Gd}^{min} \sqrt{T} = \frac{1}{C} \frac{\sqrt{2e\Gamma_1}}{\sqrt{\mathscr{D} T_D}} \frac{(R_{Gd,tot}^2 + \omega_0^2)^2}{3\gamma_e^2 B_{\perp,Gd}^2 |\omega_0^2 - R_{Gd,tot}^2|} \tag{S29}$$

When the Gd molecules are the dominant relaxation source, we can neglect the constant contribution $\Gamma_1^{const}$ and simplify the sensitivity:

$$\delta R_{Gd}^{min} \sqrt{T} \approx \frac{1}{C\sqrt{\mathscr{D} T_D}} \sqrt{\frac{2e R_{Gd,tot}}{3\gamma_e^2 B_{\perp,Gd}^2}} \frac{(R_{Gd,tot}^2 + \omega_0^2)^{3/2}}{|R_{Gd,tot}^2 - \omega_0^2|} \tag{S30}$$

We assume a photon count rate $\mathscr{D} = 10^5 s^{-1}$, which is typical for a single NV center. Considering a single NV center in a ND with diameter 20 (25) nm, with optimized Gd density which corresponds to total Gd fluctuation rate around 60.2 GHz and transverse magnetic noise $B_{\perp,Gd} \approx 146$ (104) $\mu$T, we predict a sensitivity of 6.9 (9.6) GHz in a 10 s data acquisition time. The sensitivity can be improved by measuring an ensemble of NV centers, although there might be additional noise contributions from inhomogeneities in the sensors.

## 5.2 Sensitivity to viscosity

For a fixed Gd density, all other contributions to the fluctuation rate are constant, and only the change in viscosity contributes to $\delta R_{Gd}$:

$$R_{Gd,tot} = R_{Gd,const} + \frac{k_B T}{8\pi a^3 \eta f_r} \quad \Rightarrow \quad \delta R_{Gd}^{min} = \frac{k_B T}{8\pi a^3 \eta^2 f_r} \delta\eta \tag{S31}$$

Then the sensitivity with respect to the solution viscosity is given by:

$$\delta\eta \approx \left(\frac{2e R_{Gd,tot}}{3\gamma_e^2 B_{\perp,Gd}^2 \mathscr{D} T_D T}\right)^{0.5} \frac{(R_{Gd,tot}^2 + \omega_0^2)^{3/2}}{C|(R_{Gd,tot}^2 - \omega_0^2)|} \frac{8\pi a^3 \eta^2 f_r}{k_B T} = \delta R_{Gd}^{min} \frac{\eta}{R_{rot}} \tag{S32}$$

With the estimated total fluctuation rates 6.9 (9.6) GHz mentioned above, we get the sensitivity to viscosity being 2.23 (3.11) cP using the rotational rate $R_{rot}$ in pure water case.



\end{document}